\documentclass[fleqn,usenatbib]{mnras}

\usepackage{newtxtext,newtxmath}

\usepackage[T1]{fontenc}
\usepackage{ae,aecompl}
\usepackage{float}
\usepackage{changes}

\usepackage{graphicx}	
\usepackage{amsmath}	
\usepackage{ulem}
\usepackage{amssymb}	

\title[Gravitational waves from fast-spinning white dwarfs]{Gravitational waves from fast-spinning white dwarfs}

\author[M. F. Sousa et al.]{
Manoel F. Sousa,$^{1}$\thanks{E-mail: manoel.sousa@inpe.br}
Jaziel G. Coelho,$^{1,2}$\thanks{E-mail: jazielcoelho@utfpr.edu.br} and
Jos\'e C. N. de Araujo$^{1}$\thanks{E-mail: jcarlos.dearaujo@inpe.br}
\\
$^{1}$Divis\~{a}o de Astrof\'{i}sica, Instituto Nacional de Pesquisas Espaciais, Avenida dos Astronautas 1758, S\~{a}o Jos\'{e} dos Campos, SP 12227-010, Brazil \\
$^{2}$Departamento de F\'isica, Universidade Tecnol\'ogica Federal do Paran\'a, 85884-000 Medianeira, PR, Brazil\\
}

\date{Accepted XXX. Received YYY; in original form ZZZ}

\pubyear{2019}

\begin{document}
\label{firstpage}
\pagerange{\pageref{firstpage}--\pageref{lastpage}}
\maketitle

\begin{abstract}
Two mechanisms of gravitational waves (GWs) emission in fast-spinning white dwarfs (WDs) are investigated: accretion of matter and magnetic deformation. In both cases, the GW emission is generated by an asymmetry around the rotation axis of the star. However, in the first case, the asymmetry is due to the amount of accreted matter on the magnetic poles, while in the second case it is due to the intense magnetic field. We have estimated the GW amplitude and luminosity for three binary systems that have a fast-spinning magnetized WD, namely, AE Aquarii, AR Scorpii and RX J0648.0-4418. We find that, for the first mechanism, the systems AE Aquarii and RX J0648.0-4418 can be observed by the space detectors BBO and DECIGO if they have an amount of accreted mass of $\delta m \geq 10^{-5}M_{\odot }$. For the second mechanism, the three systems studied require that the WD have a magnetic field above $\sim 10^{9}$ G to emit GWs that can be detected by BBO. We also verified that, in both mechanisms, the gravitational luminosity has an irrelevant contribution to the spindown luminosity of these three systems. Therefore, other mechanisms of energy emission are needed to explain the spindown of these objects.
\end{abstract}

\begin{keywords}
gravitational waves -- (stars:) white dwarfs -- stars: magnetic field
\end{keywords}



\section{Introduction}
\label{intro}
There is an increasing interest of the astrophysics community on highly magnetized white dwarfs (HMWDs) both from the theoretical and observational points of view. These WDs with surface magnetic fields from $10^6$~G up to $10^9$~G have  been confirmed by the recent results of the Sloan Digital Sky Survey (SDSS) \citep[see e.g.,][]{2009A&A...506.1341K,2010yCat..35061341K,2010AIPC.1273...19K,2013MNRAS.429.2934K,2015MNRAS.446.4078K}. Besides their high magnetic fields, most of them have been shown to be massive, and responsible for the high-mass peak at $1~\textrm{M}_\odot$ of the WD mass distribution; for instance: REJ 0317--853 has $M \approx 1.35~\textrm{M}_\odot$ and $B\approx (1.7$--$6.6)\times 10^8$~G \citep{1995MNRAS.277..971B,2010A&A...524A..36K}; PG 1658+441
has $M \approx 1.31~\textrm{M}_\odot$ and $B\approx 2.3\times 10^6$~G \citep{1983ApJ...264..262L,1992ApJ...394..603S};
and PG 1031+234 has the highest magnetic field $B\approx 10^9$~G \citep{1986ApJ...309..218S,2009A&A...506.1341K}. The existence of ultra-massive WDs has been revealed in several studies~\cite[see e.g.,][]{2005A&A...441..689A,2007A&A...465..249A,2013MNRAS.430...50C,2013ApJ...771L...2H,2017MNRAS.468..239C,2019A&A...625A..87C,10.1093/mnras/sty3016,2018MNRAS.480.4505J}.

Typically, WDs rotate with periods of days or even years. Recently, a pulsating WD was discovered, called AR Scorpii, that emits from X-ray to radio wavelengths, pulsing in brightness with a period of $1.97$ min~\citep{2016Natur.537..374M}.  The spindown power is an order of magnitude larger than the observed luminosity (dominated by the X-rays), which together with the absence of obvious signs of accretion suggests that AR Sco is primarily rotation-powered. Furthermore, other sources have been proposed as
candidates of WD pulsars. A specific example is AE Aquarii, the first WD pulsar identified, with a short rotation period of $P=33.08$~s~\citep{2008PASJ...60..387T}. On the other hand, the X-ray Multimirror Mission (XMM) - Newton satellite has observed a WD faster than AE Aquarii. \citet{2009Sci...325.1222M} showed that the X-ray pulsator RX J0648.0-4418 (RX J0648, hereafter) is a massive WD with mass $M=1.28M_\odot$ and radius $R = 3000$ km~\citep[see][for derived mass-radius relations for massive oxygen-neon WDs that predict this radius]{refId0,Althaus2007}, with a very fast spin period of $P = 13.2$~s, that belongs to the binary system HD 49798/RX J0648.0-4418.

On the other hand, direct observations of GWs have recently been made by LIGO and Virgo. As is well known, the first event was detected in 2015 by LIGO \citep{ABBOTT/2016}. This event, named GW150914, came from the merging of two black holes of masses $ \sim  35.6\,\rm{M}_{\odot}$ and $30.6\,\rm{M}_{\odot}$ that resulted in a black hole of mass $ \sim  63.1\,\rm{M}_{\odot}$. Thereafter, LIGO in collaboration with Virgo observed 9 more such events \citep{ABBOTT/2017a,ABBOTT/2017c, ABBOTT/2017b,Abbott_2019}. In addition, the event GW170817 reports the first detection of GWs from a binary neutron star inspiral \citep{abbott2017d}.
  
All GW detections are within a frequency band ranging from $10$ Hz to $1000$ Hz, which is the operating band of LIGO and Virgo. As is well known, there are proposed missions for lower frequencies, such as LISA \citep{AMARO/2017,cornish2018}, whose frequency band is of $(10^{-4}-0.01)$ Hz, BBO \citep{harry/2006,yagi2011} and DECIGO \citep{kawamura/2006,yagi2017} in the frequency band ranging from $0.01$ Hz to $10$ Hz. 

Different  possibilities  of  generation  of continuous GWs have already been proposed~\citep[see e.g.,][and references therein]{1996A&A...312..675B,2016ApJ...831...35D,2016JCAP...07..023D,10.1093/mnras/stx2119,2017EPJC...77..350D,2017ApJ...844..112G,Schramm/2017,2018EPJC...78..361P,2019arXiv190600774D}. More recently,~\citet{2019MNRAS.tmp.2346K} show that continuous GWs can be emitted from rotating magnetized WDs and will possibly be detected by the upcoming GW detectors such as LISA, DECIGO and BBO. Here we explore two mechanisms of gravitational radiation emission in fast-spinning magnetized WD: accretion of matter and magnetic deformation. In both cases, the GW emission is generated by asymmetry around the rotation axis of the star.

This paper is organized as follows: in Sec. \ref{sec:6} we describe the two mechanisms of GW emission by deducting the equations for gravitational amplitude and luminosity. In Sec. \ref{sec:9} we present the calculations applied to three binary systems that have a fast-spinning magnetized WD: AE Aqr, AR Sco, and RX J0648. Moreover, we discuss the results obtained in this section. Finally, in Sec. \ref{sec:13} we summarize the main conclusions and remarks.

\section{Gravitational emission mechanisms}
\label{sec:6}

WDs might generate GWs whether they are not perfectly symmetric around their rotation axes. This asymmetry can occur due to the accretion of matter \citep{CHOIYI/2000} or due to the huge dipole magnetic field that can make the star become oblate \citep{chandrasekhar1953s}. In this work, we analyze the emission of gravitational radiation from fast magnetized WDs by these two mechanisms.

\subsection{Accretion of matter}
\label{sec:7}

Here we show the luminosity and the amplitude of the GW for the case of a WD accreting matter via the magnetic poles, which do not coincide with the rotation axis of the star. In this scenario, the secondary star of the system transfers matter to the WD via an accretion column, without forming a disk, and accumulating an amount of mass on the magnetic poles~\citep[see,][and references therein]{Welsh/1998,warner/2003,Hellier/2001,lamb1987,lubow1975}. 

This configuration has been considered by \citet{CHOIYI/2000} to obtain AE Aqr's gravitational counterpart, where they assume that the spindown energy is not directly connected to any observable electromagnetic emission, but the high spindown would be caused by the accreted matter that slowly spreads over the star's surface and generates gravitational radiation. It is worth mentioning that we do not consider that all spindown is due to GWs and we have established different values for the amount of mass to calculate the observable GW amplitude.

Thus, we consider a rigid object, whose axes of symmetry are ($x_{1}$, $x_{2}$, $x_{3}$), and the corresponding main moments of inertia are $I_{1}$, $I_{2}$ and $I_{3}$, respectively. This solid rotates with angular velocity $\omega$ with respect to an axis that makes an angle $\theta$ with the $x_3$ axis. Moreover, we consider that the magnetic dipole axis is also given by the $x_3$ axis.

With this configuration and doing $I_{1} = I_{2}$, the gravitational amplitude and luminosity are given respectively by \citep[see, e.g.,][]{shapiro/2008,maggiore/2008}

\begin{equation}
  h_{ac} = \frac{4G}{c^{4}} \frac{(I_{1}-I_{3}) \omega ^{2}}{r} \sin ^{2}\theta,
  \label{AmpAC2}
\end{equation}
\par\noindent and
\begin{equation}
  L_{GW_{ac}} = -\frac{2}{5} \frac{G}{c^{5}}(I_{1}-I_{3})^{2} \omega ^{6} \sin^{2}\theta 
  \\ (16 \sin^{2}\theta + \cos^{2}\theta ),
 \label{lumAC1}
\end{equation}

\noindent where 
$r$ is the distance to the emitting source.

Now, to determine the moments of inertia $I_{1}$ and $I_{3}$, we consider that the object has deformities or an amount of mass accumulated about the $x_{3}$ axis. We reduce this system to a large sphere with two smaller spheres of matter on the $x_{3}$  axis: one at each of the poles of the larger sphere. This would be equivalent to a WD accreting matter by the two magnetic poles, where the magnetic poles do not coincide with the rotation axis of the star.
Therefore, it follows immediately that

\begin{equation}
  I_{1} = \frac{2}{5} MR^{2} + 2\delta m ~R^{2},
    \label{I1AC}
 \end{equation}

 \begin{equation}
  I_{3} = \frac{2}{5} MR^{2} + 2 \frac{2}{5} \delta m ~a^{2},
    \label{I3AC}
 \end{equation} 

\noindent where $M$ is the mass of the star, $R$ is the radius of the star, $\delta m$ is the amount of mass accumulated on one magnetic pole and $a$ is the radius of this amount.

Considering that $R \gg a$, the term $I_{1}-I_{3}$ can be expressed as follows
\begin{equation}
  I_{1}-I_{3} = 2\delta m ~R^{2}.
    \label{I1I3ACR}
 \end{equation}
By substituting this last expression into equations (\ref{AmpAC2}) and (\ref{lumAC1}),  one obtains
\begin{equation}
  h_{ac} = \frac{8G}{c^{4}} \frac{\delta m ~R^{2} \omega ^{2}}{r} \sin ^{2}\theta,
    \label{AmpAC4}
 \end{equation}
\noindent and
\begin{equation}
  L_{GW_{ac}} = -\frac{8}{5} \frac{G}{c^{5}} \delta m^{2} R^{4} \omega ^{6} \sin^{2}\theta (16 \sin^{2}\theta + \cos^{2}\theta ).
    \label{lumAC2}
 \end{equation}
Thereby, we find expressions for the gravitational luminosity and the GW amplitude for the case of a WD accumulating mass, which depends on the accreted mass, the distance to the source, the radius of the star and how fast it is rotating.

\subsection{Magnetic deformation}
\label{sec:8}

This section deals with the deformation of the WD induced by its own huge magnetic field. Let us consider that the WD is triaxial, that is, the star has asymmetries with respect to its rotation axis,  presenting a triaxial moment of inertia. In order to investigate the effect arising from the magnetic stress on the equilibrium of stars, let  us  introduce  the equatorial  ellipticity,  defined as \citep{shapiro/2008,maggiore/2008}

\begin{equation}
  \epsilon = \frac{I_{1} - I_{2}}{I_{3}}.
 \label{elipticidade}
\end{equation}

\noindent where $I_{1}$, $I_{2}$ and $I_{3}$ are main moments of inertia with respect to the ($x$, $y$, $z$) axes, respectively.

If the star rotates around the $z-$axis, then it will emit monochromatic GWs with a frequency twice the rotation frequency, $f_{rot}$, and amplitude given by \citep{shapiro/2008,maggiore/2008}

\begin{equation}
   h_{def} = \frac{16 \pi^{2} G}{c^{4}}  \frac{I_{3} f_{rot}^{2}}{r} \, \epsilon,
\label{Ampdef2}
\end{equation}

\noindent and the rotational energy of the star decreases at a rate given by \citep{shapiro/2008,maggiore/2008}

 \begin{equation}
   L_{GW_{def}} = - \frac{32}{5} \frac{G}{c^{5}} I_{3}^{2} \epsilon^{2} \omega_{rot}^{6}.
\label{Lumidef1}
\end{equation}

On the other hand, recall that the ellipticity of magnetic origin can also be written as follows~\citep{coelho2014dynamical, chandrasekhar1953s}

\begin{equation}
    \epsilon = \frac{35}{24} \frac{B_{s}^{2} R^{4}}{G M^{2}},
    \label{excentridade}
\end{equation}

\noindent where $B_{s}$ is the dipole magnetic field, $R$ and $M$ are the radius and the mass of the star, respectively.

Finally, substituting this last equation into equations (\ref{Ampdef2}) and (\ref{Lumidef1}), one immediately obtains that

\begin{equation}
   h_{def} = \frac{28 \pi^{2} }{3 c^{4}}  \frac{B_{s}^{2} R^{6} f_{rot}^{2}}{r M},
\label{Ampdef3}
\end{equation}
and
\begin{equation}
   L_{GW_{def}} = - \frac{98}{45} \frac{ B_{s}^{4} R^{12} \omega_{rot}^{6}}{c^{5} G M^{2}}.
\label{Lumidef2}
\end{equation}

Note that the two equations just above depend on the rotation frequency and the magnetic field strength.

In contrast, the GWs amplitude can also be written as a function of the variation of the star's rotation frequency $\dot{f}_{rot}$. In this case, we must consider that the whole spindown luminosity is converted into GWs. Therewith, we infer an upper limit for amplitude of GWs given by \citep{Aasi/2014}

\begin{equation}
  h_{sd} = \left ( \frac{5}{2} \frac{G}{c^{3}} \frac{I_{3} \dot{f}_{rot}}{r^{2} f_{rot}} \right )^{1/2}.
    \label{Ampsp}
\end{equation}
 
This equation must be modified to take into account that just a part of the spindown is due the GW emission. Thus, we can consider an efficiency, $\eta_{df}$, for the variation of the rotation frequency as follows

\begin{equation}
  \dot{\bar{f}}_{rot} = \eta_{df} \dot{f}_{rot},
    \label{varfreq4}
\end{equation}

\noindent such that $\dot{\bar{f}}_{rot}$ can be interpreted as the part of $\dot{f}_{rot}$ related to the GW brake. Hence, the GW amplitude can be written as follows

\begin{equation}
  h_{sd} = \left ( \eta_{df} \, \frac{5}{2} \frac{G}{c^{3}}  \frac{I_{3} \dot{f}_{rot}}{r^{2} f_{rot}} \right )^{1/2}.
    \label{Ampsp2}
 \end{equation}
 
Now, we are ready to calculate the GW amplitude and luminosity for massive fast-spinning WDs. The next section is devoted to this issue as well as the corresponding discussion of the results.

\section{Results and Discussions}
\label{sec:9}

\subsection{Accretion of matter}
\label{sec:10}

AE Aqr, is a cataclysmic variable considered peculiar because it has a WD with a very short period of rotation of $P = 33.08$ s, and a high spindown rate $\dot{P} = 5.64 \times 10^{-14}$ s s$^{-1}$ ~(see~Table~\ref{tabla: sisbin}). This value is considered high because the energy rate needed to explain the WD period variation far exceeds the quiescent luminosity observed in the ultraviolet and X-ray band or even much higher than the bolometric luminosity \citep{CHOIYI/2000}.

Several works proposed different energy emission mechanisms to explain these high spindown.  \citet{eracleous/1996} proposed a magnetic propeller model, in which the accretion flux of matter by WD is fragmented into discrete amount of mass accumulated and follows the path of the magnetic field lines. However, the fast-rotating magnetic WD moves like a fast-moving propeller expelling much of the matter from the system. Thus, the spindown energy is consumed to expel the matter. Also, \citet{CHOIYI/2000} proposed as an alternative spindown mechanism an unconventional configuration for AE Aqr. In this scenario, the WD has a magnetic dipole whose axis is misaligned with the axis of rotation. The source is accreting matter so that the accretion flux follows the field lines to the magnetic poles, and the rapid spindown is caused by the mountains of accreted matter that produces an asymmetrical deformation in the star structure generating gravitational radiation. This will be the scenario considered in this work to calculate the GW amplitude due to the accretion of matter (see section \ref{sec:7}). 

Here, we apply equation~(\ref{AmpAC4}) for the system AE Aqr, and extend to AR Sco and RX J0648 in order to calculate the GW amplitude for these objects, considering the scenario of an amount of mass accumulated on the magnetic poles. Moreover, an interesting possibility for the AR Sco is that it may represent an advanced evolutionary stage of the intermediate polar. That is, the magnetic WD may have accreted matter from its companion star by an accretion column, causing it to spin faster and faster until it has reached a very short rotation period (see~Table~\ref{tabla: sisbin}). 

\begin{table*}
\centering
\caption{Parameters of three binary systems containing a fast rotating WD: period ($P$), spindown ($\dot{P}$), adopted WD mass ($M$), radius ($R$) and distance from the system to Earth ($r$).}
\begin{tabular}{lccccc}
\hline
 \textbf{Systems} & \begin{tabular}[c]{@{}c@{}}$P$\\ (s)\end{tabular} & \begin{tabular}[c]{@{}c@{}}$\dot{P}$\\ ($10^{-15}$ s/s)\end{tabular} & \begin{tabular}[c]{@{}c@{}}$M$\\ ($M_{\odot}$)\end{tabular} &  \begin{tabular}[c]{@{}c@{}}$R$\\ ($10^{8}$ cm)\end{tabular} & \begin{tabular}[c]{@{}c@{}}$r $\\ (pc)\end{tabular} \\ \hline
 AE Aqr  & 33.08 & 56.4 & 0.80 & 7.0 & 100 \\ 
 AR Sco & 118.2 & 392 & 0.81 & 7.0 & 116 \\
 RX J0648 & 13.18 & 6.0 & 1.28 & 3.0 & 650 \\ \hline
\end{tabular}
\\ \citep[see][]{Patterson/1979,Jager/1994,CHOIYI/2000,Schramm/2017,mereghetti2011}

\label{tabla: sisbin}
\end{table*}

For these studies, we consider that the angle between the magnetic and rotation axes is $\theta = 30^{\circ}$. As a result the GW amplitude reads
\begin{equation}
  h_{ac} = \frac{2G}{c^{4}} \frac{\delta m R^{2} \omega ^{2}}{r}.
    \label{AmpAC3}
 \end{equation}
The above equation shows that the amplitude depends on the amount of mass accumulated; however, it is not easy to predict how much matter may have been accreted to WD and how much has been dispersed on its surface. Therefore, we assign here four values for the mountain of matter for the three analyzed systems: $\delta m =$ ($10^{-3} M_{\odot } $, $10^{-4}M_{\odot } $, $10^{-5}M_{\odot } $, $10^{-6}M_{\odot } $)  \citep[see e.g.,][for details about accretion in WDs]{Welsh/1998,warner/2003,Hellier/2001,lamb1987,lubow1975}.

\begin{figure}
  \includegraphics[width=8cm, height= 5.3cm]{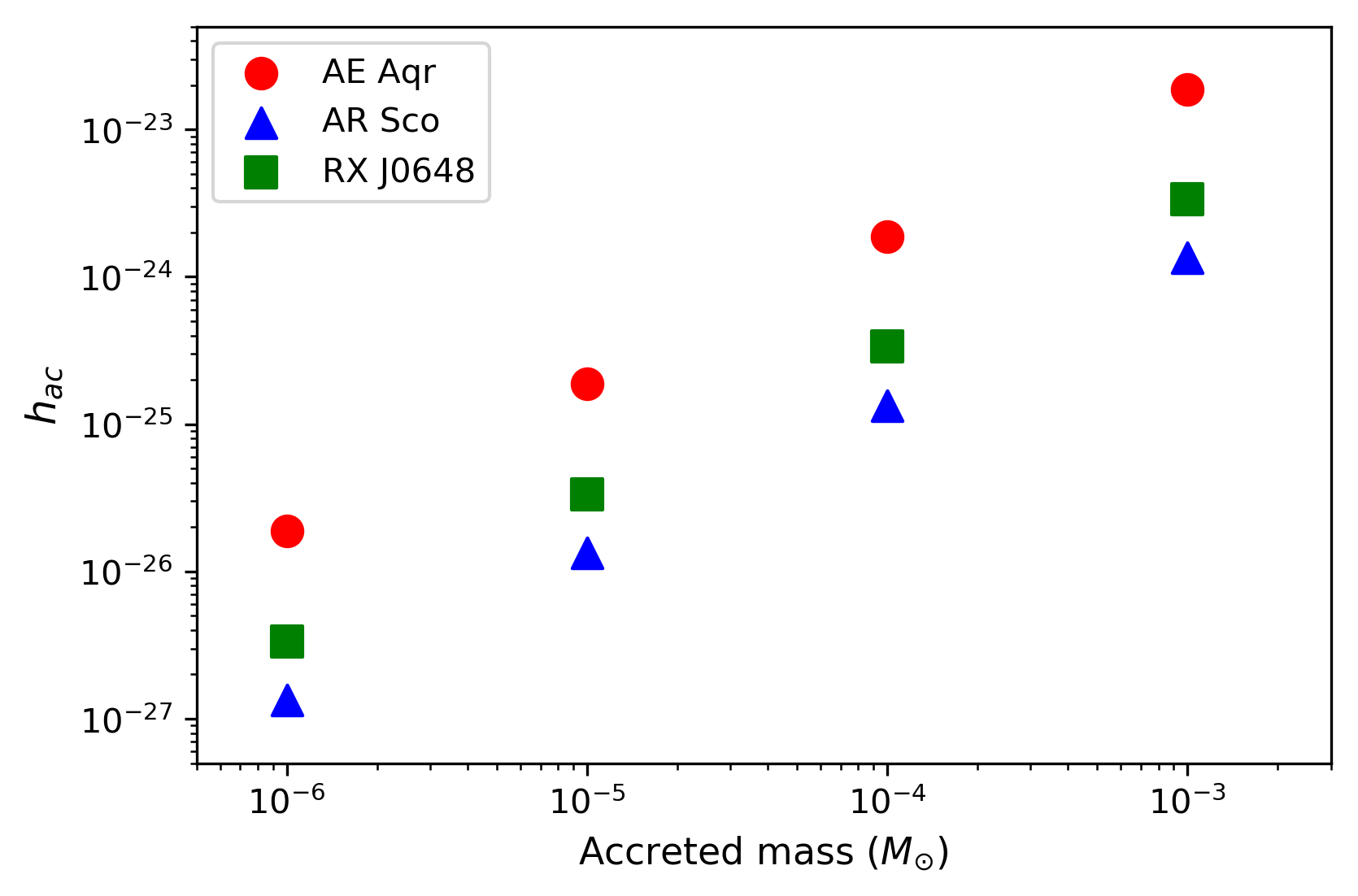}
\caption{GW amplitude as a function of accreted mass to AE Aqr, AR Sco and RX J0648}
\label{fig:1}       
\end{figure}

In addition, assuming these values for $\delta m$ and the parameters listed in Table \ref{tabla: sisbin}, we obtain $h_{ac}$ for the three systems, which are shown in Fig.~\ref{fig:1}. It is worth mentioning that for the AR Sco system that has a WD in the mass range of $0.81M_{\odot} < M_{AR} < 1.29M_{\odot}$ \citep{Schramm/2017}, we adopted the mass value of $0.81M_{\odot}$ to maximize the GW amplitude. For AE Aqr and RX J0648 systems, although mass values are not well established, we use the mass values according to \citet{CHOIYI/2000} and \citet{mereghetti2011}, respectively.

\begin{figure}
  \includegraphics[width=8cm, height= 5.5cm]{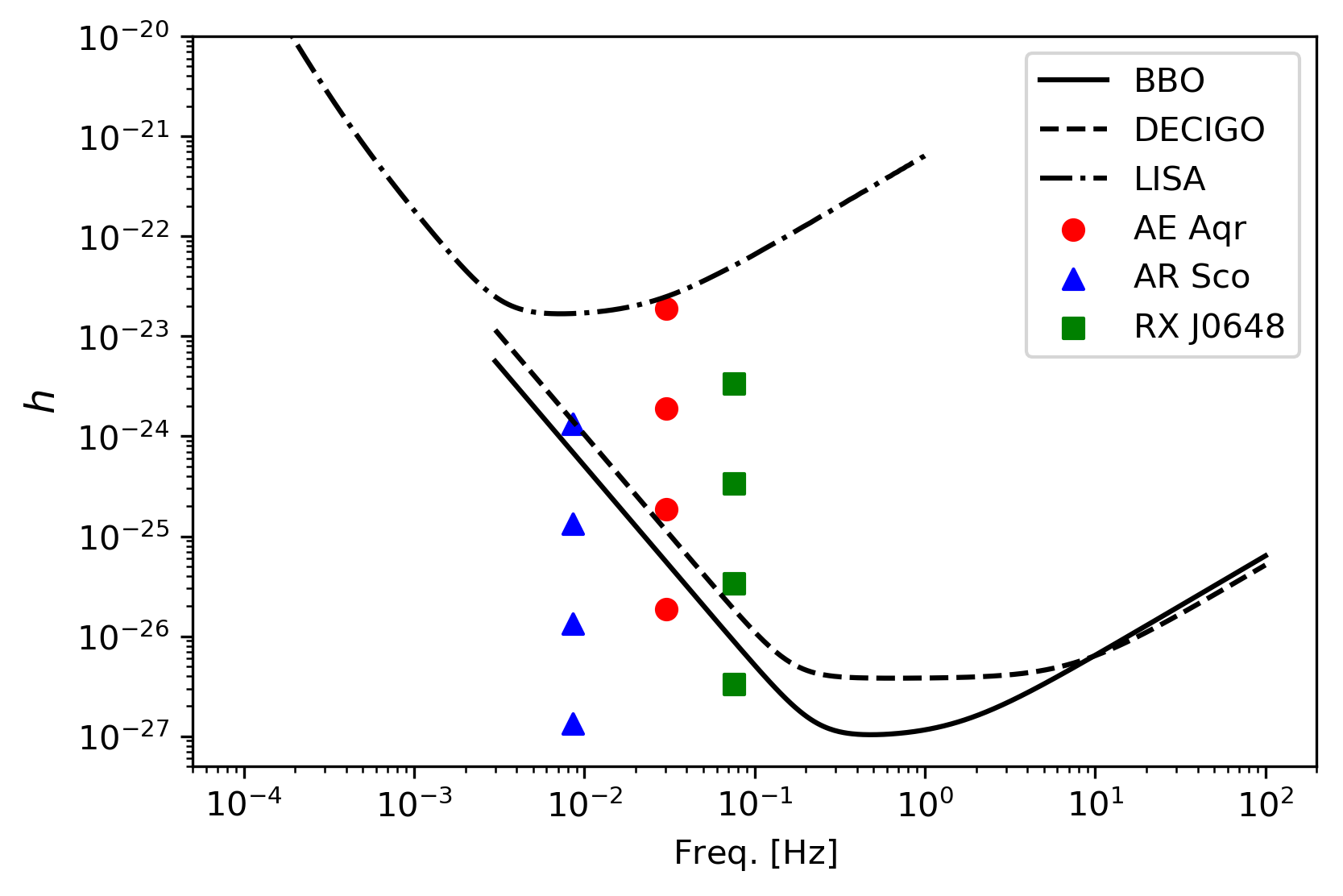}
\caption{GW amplitude for AE Aqr, AR Sco and RX J0648 for different values of mass ($10^{-3}  M_{\odot}$, $10^{-4}  M_{\odot}$, $10^{-5}  M_{\odot}$, $10^{-6}  M_{\odot}$, from top to bottom) and the sensitivity curves for LISA, BBO and DECIGO for a signal-to-noise ratio $SNR = 8$ and integration time of $T = 1$ yr.}
\label{fig:2}       
\end{figure}

At this point it is interesting to see what kind of information we can obtain from these results. Fig. \ref{fig:2} shows the GW amplitude as a function of the GW frequencies for each $ \delta m $ and the sensitivity curves of the space detectors LISA, BBO and DECIGO \citep{cornish2018,yagi2011,yagi2017}. It is worth mentioning that to plot the sensitivity curves, we use the minimum amplitude, $h_{min}$, that can be measured by the detector, for a periodic signal, for a given signal-to-noise ratio (SNR) and observation time $T$ \citep[see][for more details]{maggiore/2008}. Thereby, Fig. \ref{fig:2}, as well as Fig. \ref{fig:4}, present the GW amplitudes for the sources ($h_{ac}$ and $h_{def}$, respectively) and the sensitivity curves are set to SNR = 8 and $T = 1$ year.

Notice from Fig. \ref{fig:2}  that the systems AE Aqr and RX J0648 emit gravitational radiation with amplitudes that can be detected by BBO and DECIGO as long as $\delta m \geq 10^{-5}M_{\odot }$. For the AR Sco system, the gravitational radiation emitted by this mechanism would hardly be able to be detected by the three space instruments. This system would need to have a very high mass mountain of $ \sim 10^{-3}M_{\odot }$ to be above, for example, the sensitivity curve of the BBO detector.

\begin{table}
\centering
\caption{Spindown luminosity and accumulated mass required to explain the spindown of WD due to the emission of GWs by the mass accretion mechanism.}
\begin{tabular}{lcc}
\hline
\textbf{SYSTEMS} & \begin{tabular}[c]{@{}c@{}}$L_{sd}$\\ ($10^{33}$ erg/s)\end{tabular} & \begin{tabular}[c]{@{}c@{}}$\delta m_{sd}$\\ ($10^{-2}$ $M_{\odot }$)\end{tabular} \\ \hline
AE Aqr & 19.28 & 0.98 \\ 
AR Sco & 3.11 & 17.1 \\ 
RX J0648 & 9.53 & 0.24 \\ \hline
\end{tabular}
\label{tabla:deltamsd}
\vskip0.3cm
\end{table}

Now, we consider the efficiency of this mechanism with respect to the rotational energy rate lost by the systems. Firstly, considering that all spindown luminosity is converted in GWs, we calculate the amount of mass accumulated, $\delta m_{sd}$, required to explain the loss of rotational energy in each system, namely
\begin{equation}
  \delta m_{sd} \approx  \left ( \frac{5 c^{5} L_{sd}}{8 G R^{4} \omega^{6}} \right )^{1/2}.
    \label{deltamsd}
 \end{equation}

Table \ref{tabla:deltamsd} shows the values of this parameter for each source. We note that to explain the spindown luminosity of the star due only to the emission of GWs by the mass accretion mechanism, the WDs should have a large amount of matter at their magnetic poles. The AE Aqr and RX J0648 should have $\delta m \sim 10^{-2}$ $M_{\odot }$ and $\delta m \sim 10^{-3}$ $M_{\odot }$, respectively, while AR Sco should have an even greater $\delta m$, around $10^{-1}$ $M_{\odot }$. These values of $\delta m$ are too large, so this GW generation mechanism cannot explain the whole spindown luminosity.

We now consider the efficiency of the process ($\eta _{acr} = L_{GW_{acr}}/L_{sd}$) for the four $\delta m$'s considered above, i.e., how much of the spindown luminosity is converted to gravitational luminosity for every $\delta m$ (see Table \ref{tabla:eficie_acr}). We find that the contribution of gravitational luminosity to the spindown luminosity is irrelevant, since, for the four values of $\delta m$ adopted, the efficiency $\eta _{acr} \ll 1$, except for the source RX J0648 with a $\delta m = 10^{-3}$ $M_{\odot }$ which shows an efficiency of 17.5$\%$  (although this value of $\delta m$ can be considered too great for a WD). Therefore, other mechanisms of energy emission are needed to explain the spindown of the systems considered here.

\begin{table}
\centering
\caption{The efficiency of the generation mechanism of GWs due to the amount of mass accumulated at the WD magnetic poles for different values of $\delta m$.}
\begin{tabular}{cc}
\hline
\multicolumn{2}{c}{\textbf{AE Aquarii}} \\ \hline
\begin{tabular}[c]{@{}c@{}}$\delta m$\\ ($M_{\odot }$)\end{tabular} & \begin{tabular}[c]{@{}c@{}}$\eta _{acr}$\\ ($L_{GW_{acr}}/L_{sd}$)\end{tabular} \\ \hline
$10^{-3}$ & $1.02 \times 10^{-2}$ \\ 
$10^{-4}$ & $1.02 \times 10^{-4}$ \\ 
$10^{-5}$ & $1.02 \times 10^{-6}$ \\ 
$10^{-6}$ & $1.02 \times 10^{-8}$ \\ \hline
\end{tabular}
\hspace{0.7cm}
\begin{tabular}{cc}
\hline
\multicolumn{2}{c}{\textbf{AR Scorpii}} \\ \hline
\begin{tabular}[c]{@{}c@{}}$\delta m$\\ ($M_{\odot }$)\end{tabular} & \begin{tabular}[c]{@{}c@{}}$\eta _{acr}$\\ ($L_{GW_{acr}}/L_{sd}$)\end{tabular} \\ \hline
$10^{-3}$ & $3.41 \times 10^{-5}$ \\ 
$10^{-4}$ & $3.41 \times 10^{-7}$ \\ 
$10^{-5}$ & $3.41 \times 10^{-9}$ \\ 
$10^{-6}$ & $3.41 \times 10^{-11}$ \\ \hline
\end{tabular}
\hspace{0.7cm}
\begin{center}

\begin{tabular}{cc}
\hline
\multicolumn{2}{c}{\textbf{RX J0648}} \\ \hline
\begin{tabular}[c]{@{}c@{}}$\delta m$\\ ($M_{\odot }$)\end{tabular} & \begin{tabular}[c]{@{}c@{}}$\eta _{acr}$\\ ($L_{GW_{acr}}/L_{sd}$)\end{tabular} \\ \hline
$10^{-3}$ & $0.175$ \\ 
$10^{-4}$ & $1.75 \times 10^{-3}$ \\ 
$10^{-5}$ & $1.75 \times 10^{-5}$ \\ 
$10^{-6}$ & $1.75 \times 10^{-7}$ \\ \hline
\end{tabular}
\end{center}
\label{tabla:eficie_acr}

\end{table}

\subsection{Magnetic deformation}
\label{sec:11}

\begin{figure}
  \includegraphics[width=8cm, height= 5.3cm]{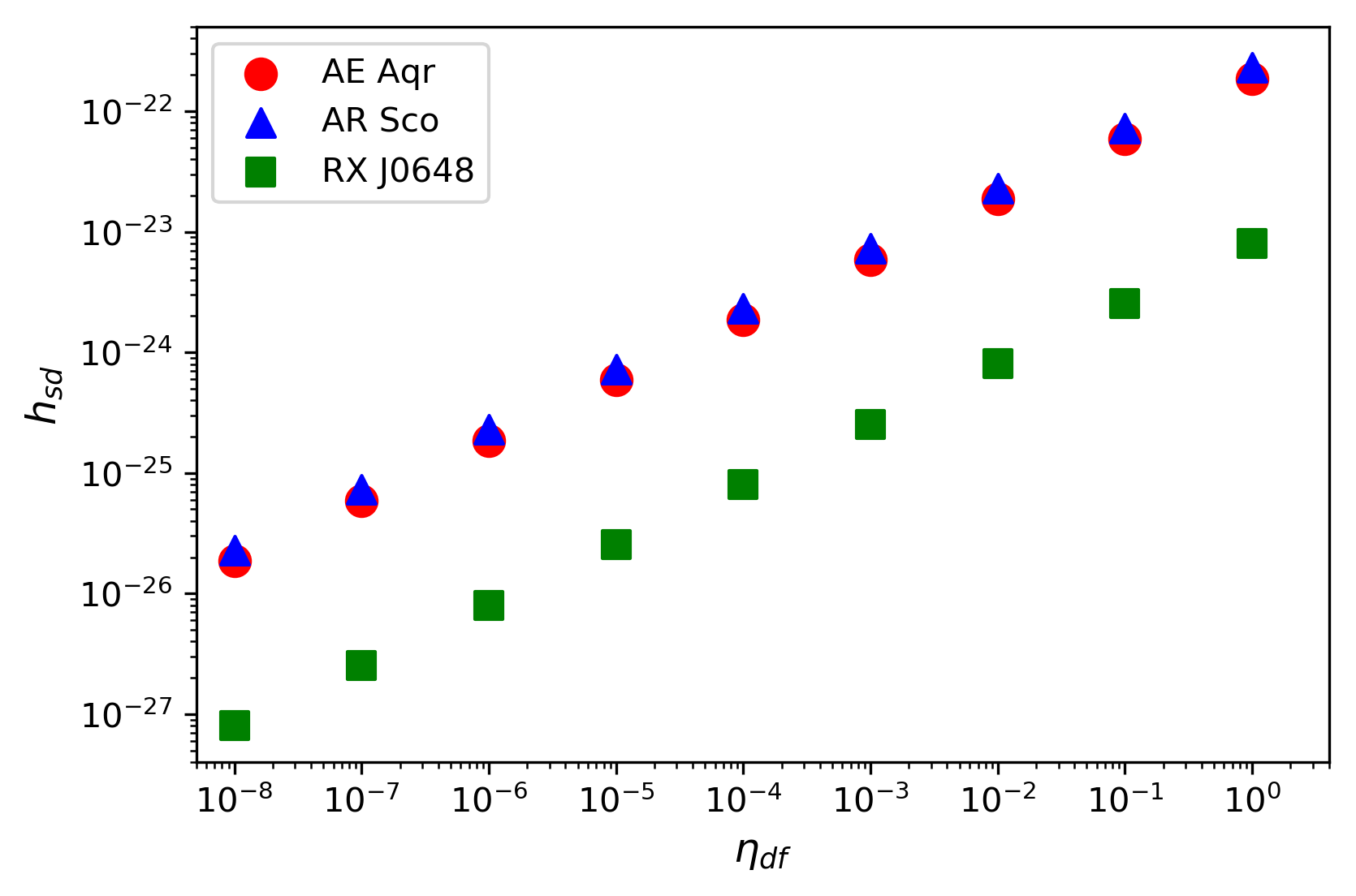}
\caption{GW amplitude for different values of efficiency ($\eta_{df} = L_{GW_{def}}/L_{sd}$) to AE Aqr, AR Sco and RX J0648}
\label{fig:3}       
\end{figure}

In this section, we consider the generation of GWs due to the deformation of the WD structure of the same binary systems (AE Aqr, AR Sco and RX J0648) caused by their own intense magnetic field. For this, we use equation (\ref{Ampsp2}) to calculate the GW amplitude as a function of the efficiency $\eta_{df} = L_{GW_{def}}/L_{sd}$. The GW amplitudes are shown in Fig. \ref{fig:3} as a function of $\eta_{df}$, where we use the parameters of Table \ref{tabla: sisbin} for all three systems.

\begin{figure}
  \includegraphics[width=8cm, height= 5.5cm]{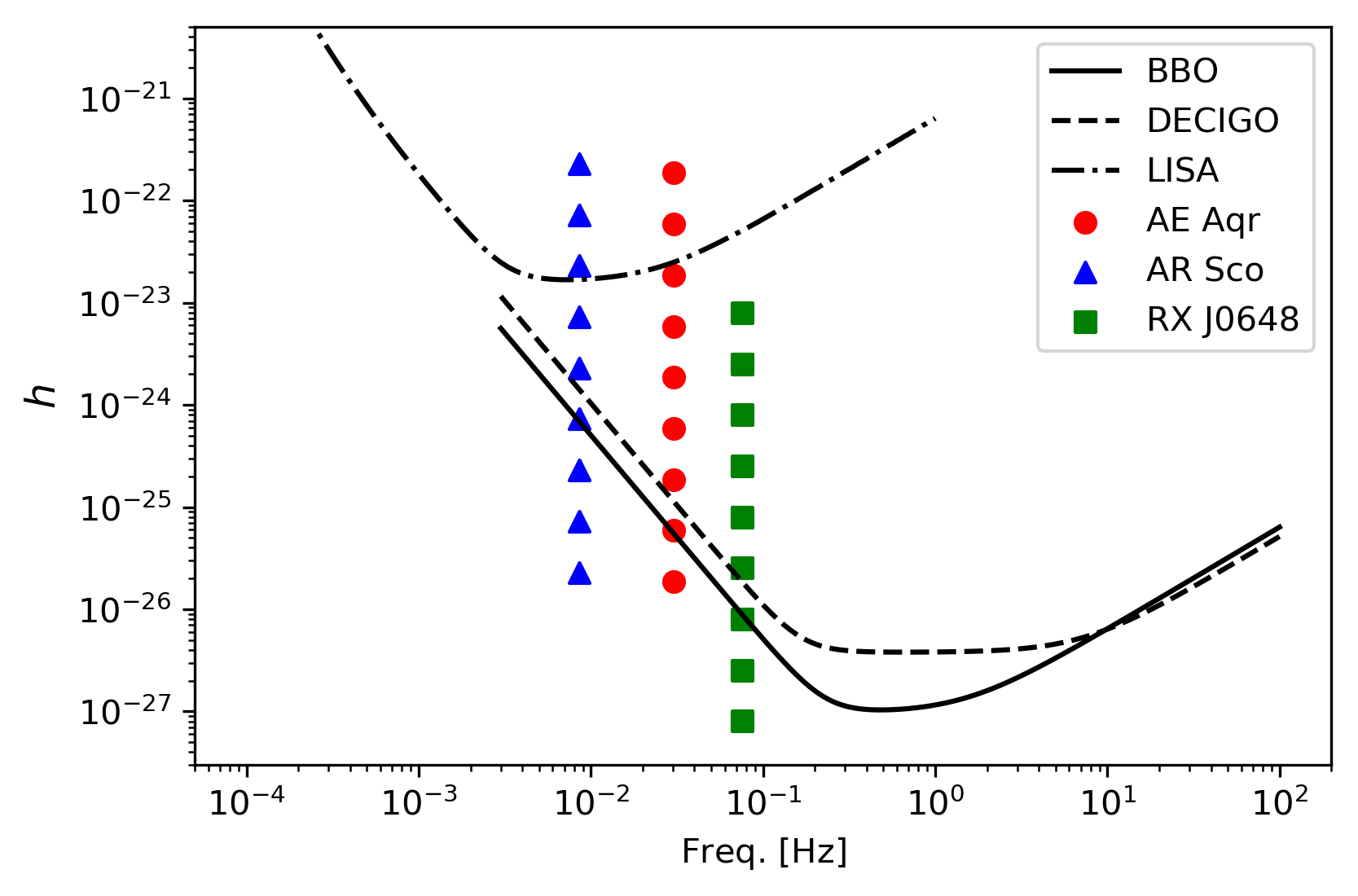}
\caption{GW amplitudes as presented in Figure \ref{fig:3} compared to the sensitivity curves of LISA, BBO and DECIGO for $SNR = 8$ and integration time of $T = 1$ year. Here, the efficiency values ($1$, $10^{-1}$, $10^{-2}$, $10^{-3}$, $10^{-4}$, $10^{-5}$, $10^{-6}$, $10^{-7}$ and $10^{-8}$) are displayed from top to bottom. }
\label{fig:4}       
\end{figure}

Fig. \ref{fig:4} shows the strain sensitivities curves for LISA, BBO and DECIGO for one year of observation time and $SNR = 8$ and the GW amplitudes as shown in Fig. \ref{fig:3}. It is worth noting that AE Aqr and AR Sco are detectable by the LISA detector, only if efficiency $\eta_{df} \geq  10^{-1}$ and $\eta_{df} \geq  10^{-2}$, respectively. On the other hand, notice that all three systems are detectable by BBO and DECIGO as long as AE Aqr has an efficiency $\eta_{df} \geq  10^{-6}$, AR Sco an efficiency $\eta_{df} \geq  10^{-4}$ and RX J0648 an efficiency $\eta_{df} \geq  10^{-5}$. Thus, even if the GWs have a small contribution to the spindown of these systems, these sources can emit GWs by the magnetic deformation mechanism with amplitudes that can be detected by the space antennas.

An interesting issue is to calculate the strength of the magnetic field needed to generate sufficient deformation to explain all the spindown luminosity. Then, using equation (\ref{Ampdef3}) we calculate the corresponding GW amplitude (see Table~\ref{tabla:eficie_def1}). Notice that the magnetic field strength obtained exceed the upper limit established by the canonical model of WD pulsars.

Also, we calculate the magnetic field strength so that these sources can be detected by BBO, which is the most sensitive instrument of the three considered in the present study. To do so, we use equation (\ref{Ampdef3}) together with the minimum efficiency for which each system is detectable by this instrument. Table \ref{tabla:eficie_def2} shows the values of these magnetic fields along with the amplitude for each system. Notice that the systems must have WDs with high magnetic fields, around $(10^{9} - 10^{10})$ G, which are about two orders of magnitude larger than the canonical model of WD pulsars.

 \begin{table}
 \centering
\caption{Magnetic field strength needed to generate enough gravitational energy to explain all the spindown luminosity.}
\begin{tabular}{lcc}
\hline
\multicolumn{3}{c}{\textbf{Efficiency $\eta _{df} = 1$}} \\ \hline
\multicolumn{1}{l}{\textbf{Systems}} & \multicolumn{1}{c}{$h_{def}$} & \multicolumn{1}{c}{$B$ (G)} \\ \hline
\multicolumn{1}{l}{AE Aqr} & \multicolumn{1}{c}{$1.85 \times 10^{-22}$} & \multicolumn{1}{c}{$8.7 \times 10^{10}$} \\
\multicolumn{1}{l}{AR Sco} & \multicolumn{1}{c}{$2.29 \times 10^{-22}$} & \multicolumn{1}{c}{$3.6 \times 10^{11}$} \\ 
\multicolumn{1}{l}{RX J0648} & \multicolumn{1}{c}{$8.02 \times 10^{-24}$} & \multicolumn{1}{c}{$2.9 \times 10^{11}$} \\ \hline
\end{tabular}
\label{tabla:eficie_def1}

\end{table}

\begin{table}
 \centering
\caption{Minimum efficiency for the sources to be measured by the BBO detector along with the amplitude of the GW and the required magnetic field strength.}

\begin{tabular}{lccc}
\hline
\multicolumn{4}{c}{\textbf{Minimum efficiency detected by BBO}} \\ \hline
\multicolumn{1}{l}{\textbf{Systems}} & \multicolumn{1}{c}{$\eta _{df}$} & \multicolumn{1}{c}{$h_{def}$} & \multicolumn{1}{c}{$B$ (G)} \\ \hline
\multicolumn{1}{l}{AE Aqr} & \multicolumn{1}{c}{$10^{-6}$} & \multicolumn{1}{c}{$1.9 \times 10^{-25}$} & \multicolumn{1}{c}{$2.8 \times 10^{9}$} \\ 
\multicolumn{1}{l}{AR Sco} & \multicolumn{1}{c}{$10^{-4}$} & \multicolumn{1}{c}{$2.3 \times 10^{-24}$} & \multicolumn{1}{c}{$3.6 \times 10^{10}$} \\ 
\multicolumn{1}{l}{RX J0648} & \multicolumn{1}{c}{$10^{-5}$} & \multicolumn{1}{c}{$2.5 \times 10^{-26}$} & \multicolumn{1}{c}{$1.6 \times 10^{10}$} \\ \hline

\end{tabular}
\label{tabla:eficie_def2}

\end{table}

\begin{table*}
\begin{center}

\caption{Elipticity ($\epsilon$), GW amplitude ($h_{def}$), GW luminosity ($L_{GW_{def}}$) and efficiency of the mechanism ($\eta _{df})$ for the upper limit of magnetic field ($B_{dip}$) of each system.}
\begin{tabular}{lccccc}
\hline
\textbf{SYSTEMS} & \begin{tabular}[c]{@{}c@{}}$B_{dip}$ \\  (G)\end{tabular} & $\epsilon$ & $h_{def}$ & \begin{tabular}[c]{@{}c@{}}$L_{GW_{def}}$ \\  (erg/s)\end{tabular} & $\eta _{df}$ \\ \hline
AE Aqr & $5.0 \times 10^{7}$ & $5.1 \times 10^{-9}$ & $6.2 \times 10^{-29}$ & $2.13 \times 10^{21}$ & $1.1 \times 10^{-13}$ \\
AR Sco & $5.0 \times 10^{8}$ & $5.3 \times 10^{-7}$ & $4.6 \times 10^{-28}$ & $1.25 \times 10^{22}$ & $4.02 \times 10^{-12}$ \\
RX J0648 & $1.0 \times 10^{8}$ & $2.8 \times 10^{-10}$ & $9.5 \times 10^{-31}$ & $1.33 \times 10^{20}$ & $1.4 \times 10^{-14}$ \\ \hline
\end{tabular}
\label{tabla:efici_def2}
\vskip0.3cm
\end{center}
\end{table*}

In addition, we can further calculate the GW amplitude by considering the upper limit values of the magnetic field strength, $B_{dip}$. Thus, knowing that AE Aqr has $B_{dip} = 5.0 \times 10^{7}$ G \citep{Isakova/2016}, AR Sco has $B_{dip} = 5.0 \times 10^{8}$ G \citep{Buckley/2017}, and RX J0648 has $B_{dip} = 1.01 \times 10^{8}$ G (inferred by the magnetic dipole model), we use equation (\ref{excentridade}) to calculate the ellipticity of the star, and equation (\ref{Ampdef3}) to calculate the GW amplitude. In addition, we compute the gravitational luminosity from equation (\ref{Lumidef2}) and the efficiency of this process with respect to the spindown luminosity.

Table \ref{tabla:efici_def2} presents the results of this study. Notice that the amplitudes of the GWs shown in this table is very small to be observed by the space detectors, since they are well below their sensitivity curves. Even for 5 years of integration time, these space instruments will not be able to detect these sources when considering those magnetic field values.

For the three binary systems investigated here, we conclude that from the magnetic deformation mechanism, the WDs require a magnetic field above $\sim 10^{9}$ G to produce GW amplitudes that can be detectable by BBO, for example. These fields are quite intense, but not unrealistic, since these WDs with surface magnetic fields from $10^6$~G up to $10^9$~G have  been confirmed by the recent results of SDSS \citep{2009A&A...506.1341K,2010yCat..35061341K,2010AIPC.1273...19K,2013MNRAS.429.2934K,2015MNRAS.446.4078K}.

It  it  worth  stressing  that  although  we  are  interested in HMWDs and most of them have been shown to be massive, it is also important that they be fast-spinning sources, in order to generate GWs in the frequency band where the space antennas are more sensitive. 
We studied here the three notable sources observed so far that would fit all these conditions (massive, fast-spinning and highly magnetized). Moreover, notice that these three sources are the fastest WDs ever observed.

Evidently, our approach could be applied to new observations of massive, fast-spinning and highly magnetized WDs. 

\section{Summary}
\label{sec:13}

After the detection of GWs from the merger events, the search for continuous GWs has been of great interest in the  scientific community. It is well known that, besides compact binaries, rapidly rotating neutron stars are promising sources of GWs which could be detected in a near future by Advanced LIGO (aLIGO) and Advanced Virgo (AdV), and also by the planned Einstein Telescope (ET). These sources generate continuous GWs whether they are not perfectly symmetric around their rotation axis, i.e., if they present some equatorial ellipticity. Undoubtedly, fast-spinning WDs are also good candidates for this purpose. Here we investigate the gravitational radiation from these uncommon WDs, which have a high rotation (a few seconds to minutes) and a huge magnetic fields ($10^{6}$ G to $\sim 10^{10}$ G), using two emission mechanisms: matter accretion and magnetic deformation. These WDs usually have a high spindown rate that is not fully explained by the electromagnetic counterpart.

Then, we study the following three binary systems: AE Aqr, AR Sco and RX J0648. Firstly, we consider the role of the aforementioned deformation due the accretion of matter in the putative generation of GWs by the HMWDs. Our calculations show that the AE Aqr and RX J0648 systems are good candidates for BBO and DECIGO if they have an amount of mass accumulated of $\delta m \geq 10^{-5}M_{\odot}$, for 1 year of integration time. AR Sco, on the other hand, is unlikely to be detected because it requires a very large amount of mass accumulated in the magnetic pole of the WD.

Secondly, regarding the magnetic deformation mechanism, we note that the three binary systems studied require that the WD has a magnetic field above $\sim 10^{9}$ G to emit gravitational radiation with amplitudes that are detectable by BBO, for example. However, these WDs are inferred to have magnetic fields with intensity around two orders of magnitude smaller.

In addition, it is worth stressing that the efficiency of both mechanisms ($\eta = L_{GW}/L_{sd}$) is very small when considering the three binary systems studied. Thus, gravitational radiation has an irrelevant contribution to the spindown luminosity of these systems.\\

\section*{Acknowledgements}
M.F.S. thanks CAPES for the financial support. J.G.C. is likewise grateful to the support of CNPq (421265/2018-3 and 305369/2018-0). J.C.N.A. thanks FAPESP (2013/26258-4) and CNPq (307217/2016-7) for partial financial support. Last but not least, we thank the referee for the suggestions and criticisms that helped to improve the manuscript.





\bibliographystyle{mnras}
\bibliography{references} 

\begin{thebibliography}{}
\makeatletter
\relax
\def\mn@urlcharsother{\let\do\@makeother \do\$\do\&\do\#\do\^\do\_\do\%\do\~}
\def\mn@doi{\begingroup\mn@urlcharsother \@ifnextchar [ {\mn@doi@}
  {\mn@doi@[]}}
\def\mn@doi@[#1]#2{\def\@tempa{#1}\ifx\@tempa\@empty \href
  {http://dx.doi.org/#2} {doi:#2}\else \href {http://dx.doi.org/#2} {#1}\fi
  \endgroup}
\def\mn@eprint#1#2{\mn@eprint@#1:#2::\@nil}
\def\mn@eprint@arXiv#1{\href {http://arxiv.org/abs/#1} {{\tt arXiv:#1}}}
\def\mn@eprint@dblp#1{\href {http://dblp.uni-trier.de/rec/bibtex/#1.xml}
  {dblp:#1}}
\def\mn@eprint@#1:#2:#3:#4\@nil{\def\@tempa {#1}\def\@tempb {#2}\def\@tempc
  {#3}\ifx \@tempc \@empty \let \@tempc \@tempb \let \@tempb \@tempa \fi \ifx
  \@tempb \@empty \def\@tempb {arXiv}\fi \@ifundefined
  {mn@eprint@\@tempb}{\@tempb:\@tempc}{\expandafter \expandafter \csname
  mn@eprint@\@tempb\endcsname \expandafter{\@tempc}}}

\bibitem[\protect\citeauthoryear{Aasi et~al.}{Aasi et~al.}{2014}]{Aasi/2014}
Aasi J.,  et~al., 2014, The Astrophysical Journal, 785, 119

\bibitem[\protect\citeauthoryear{Abbott et~al.}{Abbott
  et~al.}{2016}]{ABBOTT/2016}
Abbott B.~P.,  et~al., 2016, \mn@doi [Phys. Rev. Lett.]
  {10.1103/PhysRevLett.116.061102}, 116, 061102

\bibitem[\protect\citeauthoryear{Abbott et~al.,}{Abbott
  et~al.}{2017a}]{ABBOTT/2017a}
Abbott B.~P.,  et~al., 2017a, \mn@doi [Phys. Rev. Lett.]
  {10.1103/PhysRevLett.118.221101}, 118, 221101

\bibitem[\protect\citeauthoryear{Abbott et~al.,}{Abbott
  et~al.}{2017b}]{ABBOTT/2017c}
Abbott B.~P.,  et~al., 2017b, \mn@doi [Phys. Rev. Lett.]
  {10.1103/PhysRevLett.119.141101}, 119, 141101

\bibitem[\protect\citeauthoryear{Abbott et~al.,}{Abbott
  et~al.}{2017c}]{abbott2017d}
Abbott B.~P.,  et~al., 2017c, \mn@doi [Phys. Rev. Lett.]
  {10.1103/PhysRevLett.119.161101}, 119, 161101

\bibitem[\protect\citeauthoryear{Abbott et~al.,}{Abbott
  et~al.}{2017d}]{ABBOTT/2017b}
Abbott B.~P.,  et~al., 2017d, \mn@doi [The Astrophysical Journal]
  {10.3847/2041-8213/aa9f0c}, 851, L35

\bibitem[\protect\citeauthoryear{Abbott, Abbott, Abbott, Abraham, Acernese,
  Ackley  et~al.}{Abbott et~al.}{2019}]{Abbott_2019}
Abbott B.~P.,  Abbott R.,  Abbott T.~D.,  Abraham S.,  Acernese F.,  Ackley K.,
    et~al., 2019, \mn@doi [The Astrophysical Journal]
  {10.3847/2041-8213/ab3800}, 882, L24

\bibitem[\protect\citeauthoryear{{Althaus, L. G.}, {Garc\'{\i}a-Berro, E.},
  {Isern, J.}  \& {C\'orsico, A. H.}}{{Althaus, L. G.} et~al.}{2005}]{refId0}
{Althaus, L. G.} {Garc\'{\i}a-Berro, E.} {Isern, J.}  {C\'orsico, A. H.} 2005,
  \mn@doi [A\&A] {10.1051/0004-6361:20052996}, 441, 689

\bibitem[\protect\citeauthoryear{{Althaus, L. G.}, {Garc\'{\i}a-Berro, E.},
  {Isern, J.}, {C\'orsico, A. H.}  \& {Rohrmann, R. D.}}{{Althaus, L. G.}
  et~al.}{2007}]{Althaus2007}
{Althaus, L. G.} {Garc\'{\i}a-Berro, E.} {Isern, J.} {C\'orsico, A. H.}
  {Rohrmann, R. D.} 2007, \mn@doi [A\&A] {10.1051/0004-6361:20066059}, 465, 249

\bibitem[\protect\citeauthoryear{{Althaus}, {Garc{\'\i}a-Berro}, {Isern}  \&
  {C{\'o}rsico}}{{Althaus} et~al.}{2005}]{2005A&A...441..689A}
{Althaus} L.~G.,  {Garc{\'\i}a-Berro} E.,  {Isern} J.,   {C{\'o}rsico} A.~H.,
  2005, \mn@doi [\aap] {10.1051/0004-6361:20052996}, \href
  {https://ui.adsabs.harvard.edu/abs/2005A&A...441..689A} {441, 689}

\bibitem[\protect\citeauthoryear{{Althaus}, {Garc{\'\i}a-Berro}, {Isern},
  {C{\'o}rsico}  \& {Rohrmann}}{{Althaus} et~al.}{2007}]{2007A&A...465..249A}
{Althaus} L.~G.,  {Garc{\'\i}a-Berro} E.,  {Isern} J.,  {C{\'o}rsico} A.~H.,
  {Rohrmann} R.~D.,  2007, \mn@doi [\aap] {10.1051/0004-6361:20066059}, \href
  {https://ui.adsabs.harvard.edu/abs/2007A&A...465..249A} {465, 249}

\bibitem[\protect\citeauthoryear{Amaro-Seoane et~al.}{Amaro-Seoane
  et~al.}{2017}]{AMARO/2017}
Amaro-Seoane P.,  et~al., 2017, preprint (\mn@eprint {arXiv} {1702.00786})

\bibitem[\protect\citeauthoryear{{Barstow}, {Jordan}, {O'Donoghue}, {Burleigh},
  {Napiwotzki}  \& {Harrop-Allin}}{{Barstow}
  et~al.}{1995}]{1995MNRAS.277..971B}
{Barstow} M.~A.,  {Jordan} S.,  {O'Donoghue} D.,  {Burleigh} M.~R.,
  {Napiwotzki} R.,   {Harrop-Allin} M.~K.,  1995, \mnras, \href
  {http://adsabs.harvard.edu/abs/1995MNRAS.277..971B} {277, 971}

\bibitem[\protect\citeauthoryear{{Bonazzola} \& {Gourgoulhon}}{{Bonazzola} \&
  {Gourgoulhon}}{1996}]{1996A&A...312..675B}
{Bonazzola} S.,  {Gourgoulhon} E.,  1996, \aap, \href
  {https://ui.adsabs.harvard.edu/abs/1996A%26A...312..675B} {312, 675}

\bibitem[\protect\citeauthoryear{Buckley, Meintjes, Potter, Marsh  \&
  G{\"a}nsicke}{Buckley et~al.}{2017}]{Buckley/2017}
Buckley D.~A.~H.,  Meintjes P.~J.,  Potter S.~B.,  Marsh T.~R.,   G{\"a}nsicke
  B.~T.,  2017, Nature Astronomy, 1, 29

\bibitem[\protect\citeauthoryear{{Camisassa} et~al.,}{{Camisassa}
  et~al.}{2019}]{2019A&A...625A..87C}
{Camisassa} M.~E.,  et~al., 2019, \mn@doi [\aap] {10.1051/0004-6361/201833822},
  \href {https://ui.adsabs.harvard.edu/abs/2019A&A...625A..87C} {625, A87}

\bibitem[\protect\citeauthoryear{{Castanheira}, {Kepler}, {Kleinman}, {Nitta}
  \& {Fraga}}{{Castanheira} et~al.}{2013}]{2013MNRAS.430...50C}
{Castanheira} B.~G.,  {Kepler} S.~O.,  {Kleinman} S.~J.,  {Nitta} A.,   {Fraga}
  L.,  2013, \mn@doi [\mnras] {10.1093/mnras/sts474}, \href
  {https://ui.adsabs.harvard.edu/abs/2013MNRAS.430...50C} {430, 50}

\bibitem[\protect\citeauthoryear{Chandrasekhar \& Fermi}{Chandrasekhar \&
  Fermi}{1953}]{chandrasekhar1953s}
Chandrasekhar S.,  Fermi E.,  1953, The Astrophysical Journal, 118, 116

\bibitem[\protect\citeauthoryear{Choi \& Yi}{Choi \& Yi}{2000}]{CHOIYI/2000}
Choi C.-S.,  Yi I.,  2000, The Astrophysical Journal, 538, 862

\bibitem[\protect\citeauthoryear{Coelho, Marinho, Malheiro, Negreiros,
  C{\'a}ceres, Rueda  \& Ruffini}{Coelho et~al.}{2014}]{coelho2014dynamical}
Coelho J.,  Marinho R.,  Malheiro M.,  Negreiros R.,  C{\'a}ceres D.,  Rueda
  J.,   Ruffini R.,  2014, The Astrophysical Journal, 794, 86

\bibitem[\protect\citeauthoryear{{Curd}, {Gianninas}, {Bell}, {Kilic},
  {Romero}, {Allende Prieto}, {Winget}  \& {Winget}}{{Curd}
  et~al.}{2017}]{2017MNRAS.468..239C}
{Curd} B.,  {Gianninas} A.,  {Bell} K.~J.,  {Kilic} M.,  {Romero} A.~D.,
  {Allende Prieto} C.,  {Winget} D.~E.,   {Winget} K.~I.,  2017, \mn@doi
  [\mnras] {10.1093/mnras/stx320}, \href
  {https://ui.adsabs.harvard.edu/abs/2017MNRAS.468..239C} {468, 239}

\bibitem[\protect\citeauthoryear{{De Araujo}, {Coelho}  \& {Costa}}{{De Araujo}
  et~al.}{2016a}]{2016ApJ...831...35D}
{De Araujo} J. C.~N.,  {Coelho} J.~G.,   {Costa} C.~A.,  2016a, \mn@doi [\apj]
  {10.3847/0004-637X/831/1/35}, \href
  {https://ui.adsabs.harvard.edu/abs/2016ApJ...831...35D} {831, 35}

\bibitem[\protect\citeauthoryear{{De Araujo}, {Coelho}  \& {Costa}}{{De Araujo}
  et~al.}{2016b}]{2016JCAP...07..023D}
{De Araujo} J. C.~N.,  {Coelho} J.~G.,   {Costa} C.~A.,  2016b, \mn@doi [\jcap]
  {10.1088/1475-7516/2016/07/023}, \href
  {https://ui.adsabs.harvard.edu/abs/2016JCAP...07..023D} {2016, 023}

\bibitem[\protect\citeauthoryear{{De Araujo}, {Coelho}  \& {Costa}}{{De Araujo}
  et~al.}{2017}]{2017EPJC...77..350D}
{De Araujo} J. C.~N.,  {Coelho} J.~G.,   {Costa} C.~A.,  2017, \mn@doi
  [European Physical Journal C] {10.1140/epjc/s10052-017-4925-3}, \href
  {https://ui.adsabs.harvard.edu/abs/2017EPJC...77..350D} {77, 350}

\bibitem[\protect\citeauthoryear{{De Araujo}, {Coelho}, {Ladislau}  \&
  {Costa}}{{De Araujo} et~al.}{2019}]{2019arXiv190600774D}
{De Araujo} J. C.~N.,  {Coelho} J.~G.,  {Ladislau} S.~M.,   {Costa} C.~A.,
  2019, arXiv e-prints, \href
  {https://ui.adsabs.harvard.edu/abs/2019arXiv190600774D} {p. arXiv:1906.00774}

\bibitem[\protect\citeauthoryear{De~Jager, Meintjes, O'Donoghue  \&
  Robinson}{De~Jager et~al.}{1994}]{Jager/1994}
De~Jager O.,  Meintjes P.,  O'Donoghue D.,   Robinson E.,  1994, Monthly
  Notices of the Royal Astronomical Society, 267, 577

\bibitem[\protect\citeauthoryear{Eracleous \& Horne}{Eracleous \&
  Horne}{1996}]{eracleous/1996}
Eracleous M.,  Horne K.,  1996, The Astrophysical Journal, 471, 427

\bibitem[\protect\citeauthoryear{Franzon \& Schramm}{Franzon \&
  Schramm}{2017}]{Schramm/2017}
Franzon B.,  Schramm S.,  2017, Monthly Notices of the Royal Astronomical
  Society, 467, 4484

\bibitem[\protect\citeauthoryear{{Gao}, {Cao}  \& {Zhang}}{{Gao}
  et~al.}{2017}]{2017ApJ...844..112G}
{Gao} H.,  {Cao} Z.,   {Zhang} B.,  2017, \mn@doi [\apj]
  {10.3847/1538-4357/aa7d00}, \href
  {https://ui.adsabs.harvard.edu/abs/2017ApJ...844..112G} {844, 112}

\bibitem[\protect\citeauthoryear{Gentile~Fusillo et~al.,}{Gentile~Fusillo
  et~al.}{2018}]{10.1093/mnras/sty3016}
Gentile~Fusillo N.~P.,  et~al., 2018, \mn@doi [Monthly Notices of the Royal
  Astronomical Society] {10.1093/mnras/sty3016}, 482, 4570

\bibitem[\protect\citeauthoryear{Harry, Fritschel, Shaddock, Folkner  \&
  Phinney}{Harry et~al.}{2006}]{harry/2006}
Harry G.~M.,  Fritschel P.,  Shaddock D.~A.,  Folkner W.,   Phinney E.~S.,
  2006, Classical and Quantum Gravity, 23, 4887

\bibitem[\protect\citeauthoryear{Hellier}{Hellier}{2001}]{Hellier/2001}
Hellier C.,  2001, Cataclysmic variable stars.
Springer Praxis Books / Space Exploration, Springer

\bibitem[\protect\citeauthoryear{{Hermes}, {Kepler}, {Castanheira},
  {Gianninas}, {Winget}, {Montgomery}, {Brown}  \& {Harrold}}{{Hermes}
  et~al.}{2013}]{2013ApJ...771L...2H}
{Hermes} J.~J.,  {Kepler} S.~O.,  {Castanheira} B.~G.,  {Gianninas} A.,
  {Winget} D.~E.,  {Montgomery} M.~H.,  {Brown} W.~R.,   {Harrold} S.~T.,
  2013, \mn@doi [\apjl] {10.1088/2041-8205/771/1/L2}, \href
  {https://ui.adsabs.harvard.edu/abs/2013ApJ...771L...2H} {771, L2}

\bibitem[\protect\citeauthoryear{Isakova, Ikhsanov, Zhilkin, Bisikalo  \&
  Beskrovnaya}{Isakova et~al.}{2016}]{Isakova/2016}
Isakova P.~B.,  Ikhsanov N.~R.,  Zhilkin A.~G.,  Bisikalo D.~V.,   Beskrovnaya
  N.~G.,  2016, Astronomy Reports, 60, 498

\bibitem[\protect\citeauthoryear{{Jim{\'e}nez-Esteban}, {Torres},
  {Rebassa-Mansergas}, {Skorobogatov}, {Solano}, {Cantero}  \&
  {Rodrigo}}{{Jim{\'e}nez-Esteban} et~al.}{2018}]{2018MNRAS.480.4505J}
{Jim{\'e}nez-Esteban} F.~M.,  {Torres} S.,  {Rebassa-Mansergas} A.,
  {Skorobogatov} G.,  {Solano} E.,  {Cantero} C.,   {Rodrigo} C.,  2018,
  \mn@doi [\mnras] {10.1093/mnras/sty2120}, \href
  {https://ui.adsabs.harvard.edu/abs/2018MNRAS.480.4505J} {480, 4505}

\bibitem[\protect\citeauthoryear{{Kalita} \& {Mukhopadhyay}}{{Kalita} \&
  {Mukhopadhyay}}{2019}]{2019MNRAS.tmp.2346K}
{Kalita} S.,  {Mukhopadhyay} B.,  2019, \mn@doi [\mnras]
  {10.1093/mnras/stz2734}, \href
  {https://ui.adsabs.harvard.edu/abs/2019MNRAS.tmp.2346K} {p.~2346}

\bibitem[\protect\citeauthoryear{Kawamura et~al.}{Kawamura
  et~al.}{2006}]{kawamura/2006}
Kawamura S.,  et~al., 2006, Classical and Quantum Gravity, 23, S125

\bibitem[\protect\citeauthoryear{{Kepler}, {Kleinman}, {Pelisoli}, {Pe{\c
  c}anha}, {Diaz}, {Koester}, {Castanheira}  \& {Nitta}}{{Kepler}
  et~al.}{2010}]{2010AIPC.1273...19K}
{Kepler} S.~O.,  {Kleinman} S.~J.,  {Pelisoli} I.,  {Pe{\c c}anha} V.,  {Diaz}
  M.,  {Koester} D.,  {Castanheira} B.~G.,   {Nitta} A.,  2010, in {Werner} K.,
   {Rauch} T.,  eds,  American Institute of Physics Conference Series Vol.
  1273, American Institute of Physics Conference Series. pp 19--24,
  \mn@doi{10.1063/1.3527803}

\bibitem[\protect\citeauthoryear{{Kepler} et~al.,}{{Kepler}
  et~al.}{2013}]{2013MNRAS.429.2934K}
{Kepler} S.~O.,  et~al., 2013, \mn@doi [\mnras] {10.1093/mnras/sts522}, \href
  {http://adsabs.harvard.edu/abs/2013MNRAS.429.2934K} {429, 2934}

\bibitem[\protect\citeauthoryear{{Kepler} et~al.,}{{Kepler}
  et~al.}{2015}]{2015MNRAS.446.4078K}
{Kepler} S.~O.,  et~al., 2015, \mn@doi [\mnras] {10.1093/mnras/stu2388}, \href
  {http://adsabs.harvard.edu/abs/2015MNRAS.446.4078K} {446, 4078}

\bibitem[\protect\citeauthoryear{{Kuelebi}, {Jordan}, {Euchner}, {Gaensicke}
  \& {Hirsch}}{{Kuelebi} et~al.}{2010}]{2010yCat..35061341K}
{Kuelebi} B.,  {Jordan} S.,  {Euchner} F.,  {Gaensicke} B.~T.,   {Hirsch} H.,
  2010, VizieR Online Data Catalog, \href
  {http://adsabs.harvard.edu/abs/2010yCat..35061341K} {350, 61341}

\bibitem[\protect\citeauthoryear{{K{\"u}lebi}, {Jordan}, {Euchner},
  {G{\"a}nsicke}  \& {Hirsch}}{{K{\"u}lebi} et~al.}{2009}]{2009A&A...506.1341K}
{K{\"u}lebi} B.,  {Jordan} S.,  {Euchner} F.,  {G{\"a}nsicke} B.~T.,   {Hirsch}
  H.,  2009, \mn@doi [\aap] {10.1051/0004-6361/200912570}, \href
  {http://adsabs.harvard.edu/abs/2009A%26A...506.1341K} {506, 1341}

\bibitem[\protect\citeauthoryear{{K{\"u}lebi}, {Jordan}, {Nelan}, {Bastian}  \&
  {Altmann}}{{K{\"u}lebi} et~al.}{2010}]{2010A&A...524A..36K}
{K{\"u}lebi} B.,  {Jordan} S.,  {Nelan} E.,  {Bastian} U.,   {Altmann} M.,
  2010, \mn@doi [\aap] {10.1051/0004-6361/201015237}, \href
  {http://adsabs.harvard.edu/abs/2010A%26A...524A..36K} {524, A36}

\bibitem[\protect\citeauthoryear{Lamb \& Melia}{Lamb \& Melia}{1987}]{lamb1987}
Lamb D.~Q.,  Melia F.,  1987, in International Astronomical Union Colloquium.
  pp 511--547

\bibitem[\protect\citeauthoryear{{Liebert}, {Schmidt}, {Green}, {Stockman}  \&
  {McGraw}}{{Liebert} et~al.}{1983}]{1983ApJ...264..262L}
{Liebert} J.,  {Schmidt} G.~D.,  {Green} R.~F.,  {Stockman} H.~S.,   {McGraw}
  J.~T.,  1983, \mn@doi [\apj] {10.1086/160593}, \href
  {http://adsabs.harvard.edu/abs/1983ApJ...264..262L} {264, 262}

\bibitem[\protect\citeauthoryear{Lubow \& Shu}{Lubow \& Shu}{1975}]{lubow1975}
Lubow S.,  Shu F.,  1975, The Astrophysical Journal, 198, 383

\bibitem[\protect\citeauthoryear{Maggiore}{Maggiore}{2008}]{maggiore/2008}
Maggiore M.,  2008, Gravitational waves: volume 1: theory and experiments.
OUP Oxford

\bibitem[\protect\citeauthoryear{{Marsh} et~al.,}{{Marsh}
  et~al.}{2016}]{2016Natur.537..374M}
{Marsh} T.~R.,  et~al., 2016, \mn@doi [\nat] {10.1038/nature18620}, \href
  {http://adsabs.harvard.edu/abs/2016Natur.537..374M} {537, 374}

\bibitem[\protect\citeauthoryear{{Mereghetti}, {Tiengo}, {Esposito}, {La
  Palombara}, {Israel}  \& {Stella}}{{Mereghetti}
  et~al.}{2009}]{2009Sci...325.1222M}
{Mereghetti} S.,  {Tiengo} A.,  {Esposito} P.,  {La Palombara} N.,  {Israel}
  G.~L.,   {Stella} L.,  2009, \mn@doi [Science] {10.1126/science.1176252},
  \href {http://adsabs.harvard.edu/abs/2009Sci...325.1222M} {325, 1222}

\bibitem[\protect\citeauthoryear{Mereghetti, La~Palombara, Tiengo, Pizzolato,
  Esposito, Woudt, Israel  \& Stella}{Mereghetti et~al.}{2011}]{mereghetti2011}
Mereghetti S.,  La~Palombara N.,  Tiengo A.,  Pizzolato F.,  Esposito P.,
  Woudt P.,  Israel G.,   Stella L.,  2011, The Astrophysical Journal, 737, 51

\bibitem[\protect\citeauthoryear{Mukhopadhyay, Rao  \& Bhatia}{Mukhopadhyay
  et~al.}{2017}]{10.1093/mnras/stx2119}
Mukhopadhyay B.,  Rao A.~R.,   Bhatia T.~S.,  2017, \mn@doi [Monthly Notices of
  the Royal Astronomical Society] {10.1093/mnras/stx2119}, 472, 3564

\bibitem[\protect\citeauthoryear{Patterson}{Patterson}{1979}]{Patterson/1979}
Patterson J.,  1979, The Astrophysical Journal, 234, 978

\bibitem[\protect\citeauthoryear{{Pereira}, {Coelho}  \& {de Lima}}{{Pereira}
  et~al.}{2018}]{2018EPJC...78..361P}
{Pereira} J.~P.,  {Coelho} J.~G.,   {de Lima} R. C.~R.,  2018, \mn@doi
  [European Physical Journal C] {10.1140/epjc/s10052-018-5849-2}, \href
  {https://ui.adsabs.harvard.edu/abs/2018EPJC...78..361P} {78, 361}

\bibitem[\protect\citeauthoryear{Robson, Cornish  \& Liug}{Robson
  et~al.}{2019}]{cornish2018}
Robson T.,  Cornish N.~J.,   Liug C.,  2019, Classical and Quantum Gravity, 36,
  105011

\bibitem[\protect\citeauthoryear{{Schmidt}, {West}, {Liebert}, {Green}  \&
  {Stockman}}{{Schmidt} et~al.}{1986}]{1986ApJ...309..218S}
{Schmidt} G.~D.,  {West} S.~C.,  {Liebert} J.,  {Green} R.~F.,   {Stockman}
  H.~S.,  1986, \mn@doi [\apj] {10.1086/164593}, \href
  {http://adsabs.harvard.edu/abs/1986ApJ...309..218S} {309, 218}

\bibitem[\protect\citeauthoryear{{Schmidt}, {Bergeron}, {Liebert}  \&
  {Saffer}}{{Schmidt} et~al.}{1992}]{1992ApJ...394..603S}
{Schmidt} G.~D.,  {Bergeron} P.,  {Liebert} J.,   {Saffer} R.~A.,  1992,
  \mn@doi [\apj] {10.1086/171613}, \href
  {http://adsabs.harvard.edu/abs/1992ApJ...394..603S} {394, 603}

\bibitem[\protect\citeauthoryear{Shapiro \& Teukolsky}{Shapiro \&
  Teukolsky}{1983}]{shapiro/2008}
Shapiro S.~L.,  Teukolsky S.~A.,  1983, Black holes, white dwarfs and neutron
  stars: the physics of compact objects.
John Wiley \& Sons

\bibitem[\protect\citeauthoryear{{Terada} et~al.,}{{Terada}
  et~al.}{2008}]{2008PASJ...60..387T}
{Terada} Y.,  et~al., 2008, \mn@doi [\pasj] {10.1093/pasj/60.2.387}, \href
  {http://adsabs.harvard.edu/abs/2008PASJ...60..387T} {60, 387}

\bibitem[\protect\citeauthoryear{Warner}{Warner}{2003}]{warner/2003}
Warner B.,  2003, Cataclysmic variable stars.
Cambridge Astrophysics

\bibitem[\protect\citeauthoryear{Welsh, Horne  \& Gomer}{Welsh
  et~al.}{1998}]{Welsh/1998}
Welsh W.~F.,  Horne K.,   Gomer R.,  1998, Monthly Notices of the Royal
  Astronomical Society, 298, 285

\bibitem[\protect\citeauthoryear{Yagi \& Seto}{Yagi \& Seto}{2011}]{yagi2011}
Yagi K.,  Seto N.,  2011, Physical Review D, 83, 044011

\bibitem[\protect\citeauthoryear{Yagi \& Seto}{Yagi \& Seto}{2017}]{yagi2017}
Yagi K.,  Seto N.,  2017, Physical Review D, 95

\makeatother
\end{thebibliography}


\bsp	
\label{lastpage}
\end{document}